\documentstyle[aps,preprint,prl]{revtex}
\begin{document}
\flushbottom

\widetext
\draft
\title{Exact Dirac equation calculation of ionization and pair production 
induced by ultrarelativistic heavy ions}
\author{A. J. Baltz}
\address{
Physics Department,
Brookhaven National Laboratory,
Upton, New York 11973}
\date{October 2, 1996}
\maketitle

\def\thepage{\arabic{page}}
\makeatletter
\global\@specialpagefalse
\ifnum\c@page=1
\def\@oddhead{Draft\hfill To be submitted to Phys. Rev. C}
\else
\def\@oddhead{\hfill}
\fi
\let\@evenhead\@oddhead
\def\@oddfoot{\reset@font\rm\hfill \thepage \hfill}
\let\@evenfoot\@oddfoot
\makeatother

\begin{abstract}
An exact solution of the time-dependent Dirac equation for ionization and
pair production induced by ultrarelativistic heavy ion collisions is
presented.  Exact transition probabilities, equivalent to those that would
be obtained in an untruncated basis coupled channels calculation, are
presented.  Exact bound-electron positron pair production probabilities are
calculated to be mostly smaller than those calculated with the same potential
in perturbation theory at impact parameters small enough for differences to
occur.
\end{abstract}

\makeatletter
\global\@specialpagefalse
\def\@oddhead{\hfill}
\let\@evenhead\@oddhead
\makeatother
\nopagebreak
The calculation of bound-electron positron pair production induced by
relativisitic heavy ion collisions has been a subject great interest 
recently\cite{eich}.  One motivation for this interest is the anticipated
large rates of pair production with an electron captured into a bound state
of one of the pair of fully stripped ions in a collider such as the
Brookhaven Relativistic Heavy-Ion Collider (RHIC) or
the CERN Large Hadron Collider (LHC).  The capture process will provide an
important limit on the beam lifetime since change of the charge of an ion
leads to the loss of that ion from the beam.
Early non-perturbative coupled channel calculations showed an enhancement of
some two
orders of magnitude\cite{rum} over corresponding perturbation theory
calculations at small impact parameters for Pb + Pb reactions
at relatively low relativistivc energies (e.g., $\gamma$=2.3).  These results
motivated an extensive
investigation at ultrarelativistic energies such as will occur at RHIC
($\gamma$=23 000), and the problem of non-perturbative enhancemant was shown to
present no serious obstacle to machine performance.
It was found that the enhancement
over perturbation theory systematically decreased with increasing basis size
of the coupled channels calculations
up to the largest basis size attainable, where the total non-perturbative
enhancement was found to be only of order 10\% of the total cross 
section\cite{brw4}.  It was further observed that the limited enhancement
over perturbation theory applied as well at other ultrarelativistic energies
(such as at LHC) since the probabilities at impact parameters small enough to
have non-perturbative effects are $\gamma$ independent at large
$\gamma$\cite{brw}.

In this Letter it will be shown that, in the ultrarelativistic limit,
the time-dependent Dirac equation can be solved exactly, and
one discovers that exact semi-classical probabilities of
bound-electron positron pair production are actually less than those calculated
in perturbation theory.  One also obtains exact results for single electron
ionization and finds that they are consistent with unitarity.  As a corollary,
ionization calculations can now be carried out without consideration of the
continuum final states but rather by considering the flux lost from the initial
bound state.

It has recently been shown\cite{ajb} that, in the appropriate gauge \cite{brw},
the Coulomb potential produced by an ultrarelativistic particle (such as a
heavy ion) in uniform motion takes the following form
\begin{equation}
V(\mbox{\boldmath $ \rho$},z,t)
=-\delta(z - t) \alpha Z_P(1-\alpha_z)
\ln{({\bf b}-\mbox{\boldmath $ \rho$})^2 \over b^2 }.
\end{equation}
${\bf b}$ is the impact parameter, perpendicular to the $z$--axis along which
the ion travels, $\mbox{\boldmath $\rho$}$, $z$, and $t$ are the coordinates of
the potential relative to a fixed target (or ion),
$\alpha_z$ is the Dirac matrix, $\alpha$ is the fine structure constant,
and $Z_P, v$ and $\gamma$ are the charge, velocity and $\gamma$ factor the
moving ion $(\gamma=1/\sqrt{1-v^2})$.  This is the physically relevant
ultrarelativistic potential since it was 
obtained by ignoring terms in
$ 1 / \gamma^2$\cite{ajb}\cite{brw}.
The $b^2$ in the denominator of the logarithm is
removable by a gauge transformation, and if one wished to have a potential with
the same gauge for all impact parameters one would remove it.  However, the
freedom to include or remove the extra $b^2$ will be retained for possible
computational convenience and as a minimal test of gauge invariance.

It was suggested in Ref. \cite{ajb} that the reduction of the interaction
from three dimensions to the two of Eq.(1) might make direct solution of the
time-dependent Dirac equation, without using coupled channels, a viable
alternative for the calculations of pair production induced by
ultrarelativistic heavy ions.  I point out in this Letter that,
in fact, the delta function form of Eq.(1) allows exact evaluation of the
transition amplitudes for pair production without using coupled channels.
The form that the amplitudes take is that of perturbation theory, but with
a universal effective interaction, modified from the the lowest order
interaction to exactly include coupling to all orders.

The time-dependent Dirac equation that I wish to solve is 
\begin{eqnarray}
{i\partial\Psi({\bf r},t)\over\partial t}&= \Bigl[ H_1 &\Bigr]
 \Psi({\bf r},t) \nonumber \\
&=\Bigl[H_0 &- 
V(\mbox{\boldmath $ \rho$},z,t)\Bigr]\Psi({\bf r},t) \nonumber \\
&=\Bigr[H_0 &+ \delta(z - t) \alpha Z_P(1-\alpha_z) \nonumber \\
&& \times \ln{({\bf b}-\mbox{\boldmath $ \rho$})^2 }\Bigl]\Psi({\bf r},t) ,
\end{eqnarray}
where $H_0$ is a time-independent Dirac hamiltonian, in this case of an
electron in the Coulomb field of a one of the ions (target) in its rest frame,
\begin{equation}
H_0 = \mbox{\boldmath $\alpha$}{\bf p} + \beta - \alpha Z_T /r,
\end{equation}
$\Psi({\bf r},t)$ is the exact four component time-dependent
Dirac spinor solution, and for typographical simplicity the gauge
explicitly shown is without
the $b^2$ in the denominator of the logarithm.

In the usual coupled channels approach\cite{rum} one expands the solutions of
Eq.(2) in a time-independent basis of eigenfunctions of $H_0$,
\begin{equation}
\Psi^j({\bf r},t) = \sum_k a_k^j(t) \phi_k(r) e^{-i E_k t},
\end{equation}
and then substitutes Eq.(4)
into both left and right hand sides of Eq.(2) to obtain coupled equations for
the time-dependent amplitudes, $a_k^j(t)$.  For electron-positron pair
production
$\phi_k$ includes bound electron states, continuum electron states,
and states in the negative energy
continuum.  Pair production may be represented as a transition from
an initial negative continuum state to a final bound or positive continuum
electron state.  How this scheme preserves the Pauli principle for
non-interacting electrons and how time reversal can be exploited in these
calculations has been previously discussed.\cite{rum} 

The present treatment differs from the usual coupled channels approach in that
I substitute Eq.(4) only into the left hand side of Eq.(2).  In the usual way,
I then multiply both sides by any particular state $\phi_f$, and perform the
spatial integration to obtain
\begin{eqnarray}
{d a_f^j(t) \over d t} 
&= -i e^{i E_f t} \langle \phi_f \vert & \delta(z - t) \alpha Z_P(1-\alpha_z)
\nonumber \\
& &\times
\ln{({\bf b}-\mbox{\boldmath $ \rho$})^2 } \vert \Psi^j({\bf r},t) \rangle .
\end{eqnarray}
The initial condition,
$a_f^j(t=-\infty) = \delta_{fj}$, is specified by the index $j$ and given
equivalently
\begin{equation}
\Psi^j({\bf r},t=-\infty) = \phi_j(r) e^{-i E_j t_{-\infty}}.
\end{equation}

Of course, if one knew the exact solution, $\Psi^j({\bf r},t)$, then it would
be possible simply to integrate Eq.(5) over $t$ and obtain the exact
scattering amplitudes,
$a_f^j(t=\infty)$.  But the delta function and $(1-\alpha_z)$ factor in Eq.(5)
mean that one needs only
to know $(1-\alpha_z)\Psi^j({\bf r},t)$ at $z = t$.  And it turns out that
one can obtain $(1-\alpha_z)$ times the exact solution of Eq.(2) in the region
near $z = t$ in the
following way.  First temporarily express Eq.(2) in terms of the usual light
cone coordinates
\begin{eqnarray}
x^+ &=& { 1 \over \sqrt{2} } ( t + z ) \nonumber \\
x^- &=& { 1 \over \sqrt{2} } ( t - z )
\end{eqnarray}
instead of $t$ and $z$.  Integration of $x^-$ across the $\delta$ function
then gives
\begin{equation}
(1-\alpha_z)\Psi^j(\mbox{\bf r},t)=(1-\alpha_z)e^{-i \theta(t-z) \alpha Z_P 
\ln{({\bf b}- \mbox{\boldmath $ \rho$})^2 }}
\phi_j(r) e^{-i E_j t} ,
\end{equation}
valid for $t < z$ and in the region near $t = z$ (i.e at $t = z$ and
$t = z + \epsilon$). 
Substituting Eq.(8) into Eq.(5) and integrating over $t$, I obtain
\begin{eqnarray}
a_f^j(t=\infty)=\delta_{fj} &
-i & \int_{-\infty}^{\infty} d t e^{i (E_f - E_j) t}\nonumber \\ 
&\times  \langle & \phi_f \vert
\delta(z - t) \alpha Z_P(1-\alpha_z)
\ln{({\bf b}-\mbox{\boldmath $ \rho$})^2 } \nonumber \\
&&\times \vert e^{-i \theta(t-z) \alpha Z_P 
\ln{({\bf b}-
\mbox{\boldmath $ \rho$})^2 }}\phi_j \rangle.
\end{eqnarray}
Now since by definition,
$\delta(u) = d \theta(u) / d u$,
one obtains upon carrying out the $t$ integration
\begin{eqnarray}
a_f^j(t=\infty)&=\delta_{fj} & +
\langle \phi_f \vert
(1-\alpha_z) e^{i (E_f - E_j) z} \nonumber \\ 
&& \times ( e^{-i \alpha Z_P \ln{({\bf b}- 
\mbox{\boldmath $ \rho$})^2 }} - 1 ) 
\vert \phi_j \rangle.
\end{eqnarray}
thus Eq.(9) may be equivalently expressed in the form
\begin{eqnarray}
a_f^j(t=\infty)&=\delta_{fj} & +
\int_{-\infty}^{\infty} d t e^{i (E_f - E_j) t} \langle \phi_f \vert
\delta(z - t) (1-\alpha_z) \nonumber \\ 
&&\times ( e^{-i \alpha Z_P \ln{({\bf b}- \mbox{\boldmath $ \rho$})^2 }} - 1 ) 
\vert \phi_j \rangle
\end{eqnarray}
since Eq.(10) trivially follows from it.  We make use of Eq.(11) for the
calculations since the angular momentum algebra of the computer code makes use
of the Legendre polynomial series for the $\delta$ function\cite{ajb}
\cite{brw}.

One now has a simple matrix element expression that is equivalent to the
solution of the full coupled channels problem with no truncation of basis.
The full solution of the problem, Eq.(11), is in perturbation
theory form, but with a universal effective interaction $i \delta(z - t)
(1-\alpha_z) ( e^{-i \alpha Z_P \ln{({\bf b}- \mbox{\boldmath $ \rho$})^2 }}
 - 1 )$ 
instead of the perturbation interaction $\delta(z - t)
(1-\alpha_z) \alpha Z_P \ln{({\bf b}- \mbox{\boldmath $ \rho$})^2 }$.  
The only difference between the perturbative and exact matrix
element expressions comes in the m-dependent form factors of the interaction,
where instead of the analytical Fourier transforms \cite{ajb}\cite{brw}
of the real
$\ln{({\bf b}-\mbox{\boldmath $ \rho$})^2 }$ one must substitute Fourier
transforms of the complex $ i ( e^{-i \alpha Z_P \ln{({\bf b}- 
\mbox{\boldmath $ \rho$})^2 }} - 1 ) / \alpha Z_P$ to be evaluated numerically.
These exact matrix elements exhibit time reversal symmetry because they are
in perturbation theory form with an effective potential.  
 
In the conventional coupled channels method of calculating bound-electron
positron pair production one makes use of time reversal and makes the
bound electron the initial state\cite{rum}.  Both
positive and negative electron states are coupled.  Since
the rate of excitation to negative continuum states is about three orders of
magnitude smaller than to positive excited electron states, one thereby also
calculates the ionization probability for a single bound electron.  From
Eq.(10) one may obtain in simple form the exact survival probability of an
initial state $j$
\begin{equation}
P_j (b) =  \vert \langle \phi_j \vert (1-\alpha_z)
e^{-i \alpha Z_P \ln{({\bf b}- \mbox{\boldmath $ \rho$})^2 }} 
\vert \phi_j \rangle \vert^2.
\end{equation}

In our previously reported large basis coupled channels calculations of
bound-electron positron pair production \cite{brw4} we found about an overall
non-perturbative enhancement of $7 \pm 2$ barns  over the 112 barn perturbation
theory result for Pb + Pb at RHIC\cite{brwp}.
For large contributing impact parameters we found
a negligible non-perturbative enhancement.  For the smallest impact parameters
our best truncated calculations showed non-perturbative enhancement still on
the order of a factor of two.  Parallel calculations have now been carried
out using the present exact expressions.  Results are presented in Table I.
At every impact parameter (except zero) the exact probability is smaller than
the perturbation theory result.  
Contrary to our previous result,
the exact formalism yield a small suppression rather than enhancement
for the rate of bound-electron
positron pair production due to non-perturbative effects.  For a large set of
final states (corresponding to 61\% of the perturbation theory cross section)
I find the exact evaluation yields a cross section about 3 barns {\it less}
than the perturbation theory evaluation.  The corresponding coupled channels
calculation gave about 9 barns {\it more} than perturbation theory.

A calculation of exact amplitudes for excitation of an initial electron state
into allowed final states should exhibit unitarity, which, of course, is absent
in perturbation theory.
To demonstrate that there is no apparent violation of unitarity, calculations
of single electron ionization have been performed at various impact
parameters.  Results are shown in Table II.
Because of the huge low excitation energy contribution to ionization, the
cuttoff at $\vert k \vert \leq 7$ and E $\leq 16.8 m_e c^2$ apparently covers a
relatively greater
part of the ionization cross section (about $90 \%$) than the bound-electron
positron cross section ($61 \%$).  Note that at no impact parameter does the
sum of final state bound and continuum probabilities exceed unity; the
$ 10 \%$ of the continuum electron probability missing is presumably due
to the cuttoff in energy and angular momentum, and is not inconsistent with
a smooth extrapolation of $\vert k \vert$ and E to infinity.  
To look at it another way, note that with
the present method single electron ionization cross sections can be can be
calculated without even considering continuum wave functions.  One simply
subtracts the sum of the ground state survival probability (column 2) and the
excited bound state probabilities (column 3) from unity at each impact
parameter to obtain the ionization probability at that impact parameter.

One might reasonably ask what is the physical reason that the exact
probablilities are less than those calculated in perturbation theory.
For the ionization calculations the answer is clearly unitarity: in the limit
of large $\alpha Z_P$ the sum of all perturbative probabilities for
transitions to excited states must eventually exceed unity.  An exact
calculation must maintain conservation of probability.  And although the
sum of the perturbative bound-electron positron probabilities is several
orders of magnitude
smaller, it too must violate unitarity in the strong coupling limit.

But there is another aspect of the reaction that explains the failure of
coupled channels to provide the correct sign of the correction to perturbation
theory for bound-electron positron production: the reaction is highly
adiabatic.  Figure 1 shows the time development
of the total flux in ground state electron plus continuum positron states for
a relatively small atomic impact parameter (125 fm) where the time dependent
field is relatively strong.
The time-dependent component of the field adiabatically excites and then
deexcites bound-electron positron pairs.  There is a very delicate
cancellation in the positive and negative time contributions to the
amplitudes.  The exact probability (solid line) rises and falls similarly
with the perturbative probability (dashed line) but with a smaller magnitude.
The coupled channels calculation (dot-dashed line) has the smallest maximum
of the curves
(at $t = 0$), but is the largest asymptotically.  The coupled channels
calculation was performed using rather crude wave packets for the unbound
negative and positive electrons\cite{brw4}.  The exact and perturbative
calculations
were performed using continuum wave functions (heavy lines) with the
same calculations performed using corresponding wave packets shown in the
faint lines for comparison.

In a simple test of gauge invariance the exact calculation using continuum
wave functions apparently passes, but the corresponding calculation using
wave packets is less successful.  Because the analytical Fourier series
of the perturbative potential arises from the gauge of Eq.(1), 
that gauge has been utilized in the calculations so far reported here (except,
of course, at $b = 0$).  Thus to remove the $b^2$ in the denominator of the
logarithm, one must add a corresponding scalar $\ln{b^2}$ term to the
interaction.  At large $b$ the dominant term of the Fourier series is of
dipole form, $\rho / b$, leading to the $ 1 / b^2 $ falloff of the
probabilities.  But the $\ln{b^2}$ term added by the gauge transformation
increases with $b$ while the physical dipole term decreases.  Calculations
have therefore been
performed at $b$ = 8000 fm to test gauge invariance.  With continuum wave
functions the exact result shows a change from $1.036 \times 10^{-6}$ 
to $1.062 \times 10^{-6}$ under the gauge tranformation while the 
corresponding wave packet
results goes from $1.041 \times 10^{-6}$ to $1.244 \times 10^{-6}$. These
results were calculated at a mesh size of .025 with the differences in the
two gauges dropping precipitously from those calculated at our standard
mesh of .05 in the continuum case but not in the wave packet case.  There
is no surprise here.  We
would not expect calculations done with wave packets to be exactly gauge
invariant.  The packet states are not exact eigenfunctions of the
time-independent hamiltonian $H_0$ and there is a lack completeness.
On the other
hand, using continuum wave functions, no disagreements inconsitent with
expected numerical accuracy were found between exploratory gauge transformed
calculations and the results of Tables I and II.  For example, in the strong
coupling impact parameter case of $b$ = 125 fm the ionization probability
changes by 0.2\% under the gauge transformation and the bound-electron
positron probability changes by 0.8\% under the same transformation.

Apparently some combination of basis truncation and the necessity of using
wave packets for continuum-continuum coupling provides an intractable
limitation on the coupled channels method, thereby making it incapable of
adequate evaluation of the adiabatic  cancellation.  This failure of the
coupled channels
method in the ultrarelativistic limit makes one question its utility
even at more modest relativisitic energies, where the same properties
of adiabaticity, basis truncation, and wave packets remain.

The above exact method should have other applications in the future.
The corresponding perturbation theory cross section for continuum-electron
positron pair production at RHIC is about 30,000 barns.  Non-perturbative
calculations for this process have so far been prohibitive in difficulty
because of the large number of energy and angular momentum states that are
coupled.  The present
approach now seems to make the problem tractable.  It allows the exact
cross section to be calculated for
any particular electron-positron pair in an expression decoupled from all the
other pairs.

I am indebted to J. Weneser for discussion and for previous collaborative
work that paved the way for the present results.

\vskip .5cm
This manuscript has been authored under Contract No. DE-AC02-76-CH00016 with
the U. S. Department of Energy. 

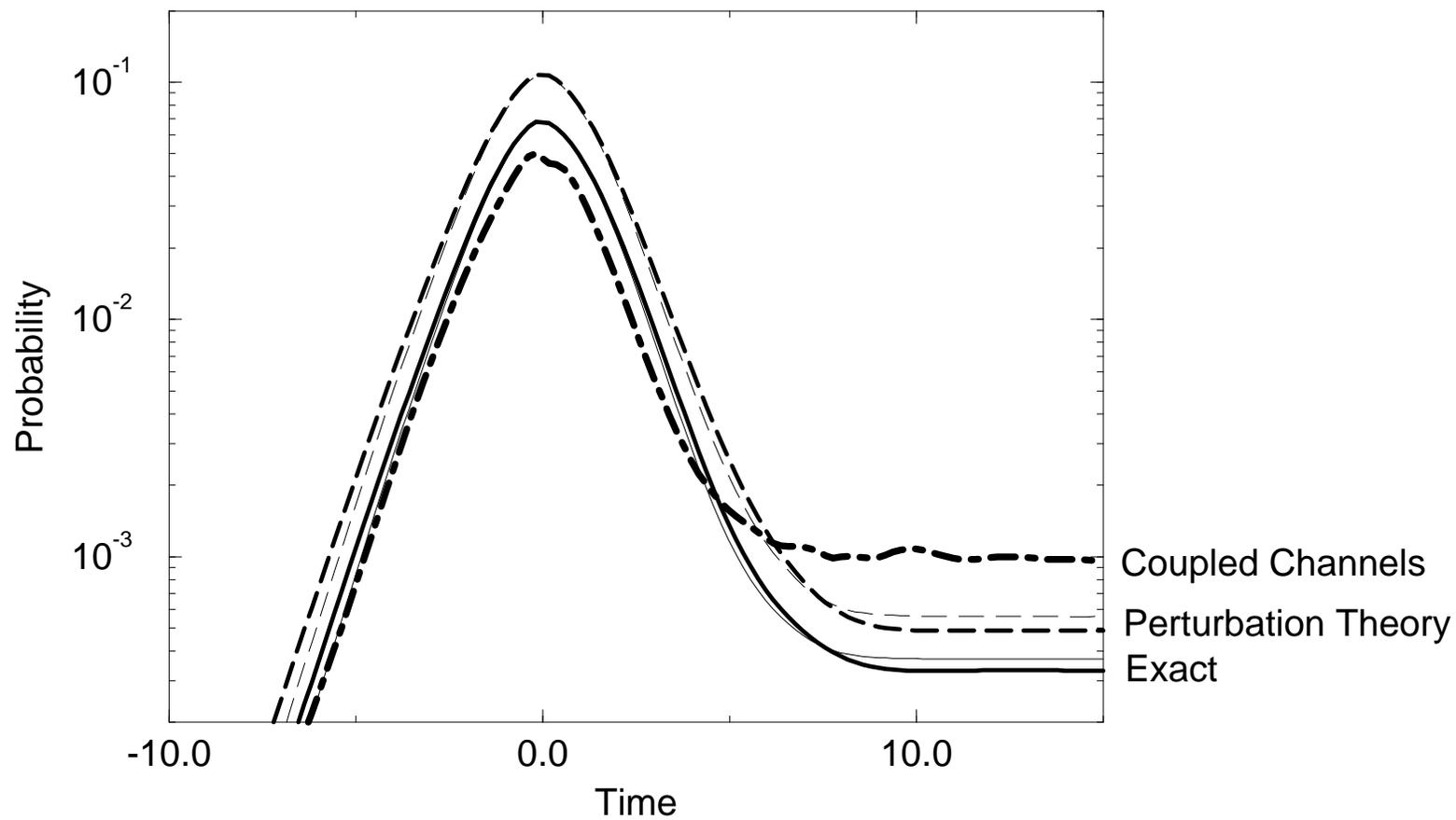
\begin{figure}
\caption[Figure 1]{Probability of excitation of a K-orbit electron plus any
continuum positron by a Pb ion impinging on a Pb ion target.  Impact parameter
is 125 fm.  Time is in natural units (386 fm.)}
\label{pair}
\end{figure}
\begin{table}
\caption[Table I]{Bound-electron positron pair production
probabilities for Pb + Pb (to be multiplied by $ 2 \times 10^{-6}$) are in the
second and third columns. The fourth and fifth columns are in barns and
represent the cross section differences from perturbation theory in the
annulus from b/$\sqrt{2}$ to $\sqrt{2}$b.}
\begin{tabular}{ccccc}
b(fm) & Exact & Perturb.& Exact Enhance. & C. C. Enhance. \\
0 & 289. & 283. & .001 & $<$ .08 \\
62.5 & 271. & 505. & $-$.08 & $<$ .22 \\
125 & 330. & 487. & $-$.21  & .53 \\
250 & 297. & 432. & $-$.73 & .73 \\
500 & 171. & 216. & $-$.98 & 1.26 \\
1000 & 61.5 & 67.7 & $-$.54 & 3.00 \\
2000 & 16.57 & 16.92 & $-$.12 & 2.30 \\
4000 & 4.144 & 4.160 & $-$.02 & .82 \\
8000 & 1.0348 & 1.0357 & $-$.005 & .20 \\
\end{tabular}
\label{tabi}
\end{table}
\begin{table}
\caption[Table II]{Ionization and Unitarity: Probabilities for Pb + Pb}
\begin{tabular}{ccccc}
b(fm) & $e^-_{gr}$ & $\sum_{bnd} e^-_{ex}$ & 
$\sum_{cont} e^-$ &
$\sum e^- $\\
0 & .428 & .100 & .457 & .985 \\
31.25 & .430 & .099 & .454 & .983 \\
62.5 & .434 & .099 & .444 & .978 \\
125 & .447 & .101 & .426 & .974 \\
250 & .488 & .104 & .381 & .974 \\
500 & .582 & .101 & .292 & .975 \\
1000 & .730 & .086 & .169 & .986 \\
2000 & .890 & .054 & .052 & .996 \\
4000 & .971 & .019 & .009 & .999 \\
8000 & .9927 & .0051 & .0020 & .9998 \\
16000 & .99818 & .00128 & .00049 & .99995 \\
\end{tabular}
\label{tabii}
\end{table}
\end{document}